\documentclass[twocolumn,showpacs,preprintnumbers,amsmath,amssymb]{revtex4}

\usepackage{graphicx}
\usepackage{dcolumn}
\usepackage{bm}

\begin{document}

\title{Vortex assisted mechanism of photon counting in
superconducting nanowire single photon detector revealed by
external magnetic field}

\author{D.Yu. Vodolazov$^{1,2}$}
\affiliation{$^1$ Institute for Physics of Microstructure, Russian
Academy of Sciences, 603950,
Nizhny Novgorod, GSP-105, Russia \\
$^2$ Lobachevsky State University of Nizhny Novgorod, 23 Gagarin
Avenue, 603950 Nizhny Novgorod, Russia}
\author{Yu.P. Korneeva$^{3}$, A.V. Semenov$^{3,4}$, A.A. Korneev$^{3,4,5}$, G.N. Goltsman$^{3,5}$}
\affiliation{$^3$ Moscow State Pedagogical University, 119991, 1 M. Pirogovskaya st., Moscow, Russia\\
$^4$ Moscow Institute of Physics and Technology, 141700, Moscow region, Dolgoprudny, 9 Institutsky lane, Russia\\
$^5$ National Research University -- Higher School of Economics, 101000, 20 Myasnitskaya st., Moscow, Russia}

\date{\today}

\begin{abstract}

We use external magnetic field to probe the detection mechanism of
superconducting nanowire single photon detector. We argue that the
hot belt model (which assumes partial suppression of the
superconducting order parameter $\Delta$ across the whole width of
the superconducting nanowire after absorption of the single
photon) does not explain observed weak field dependence of the
photon count rate (PCR) for photons with $\lambda$=450 nm and
noticeable {\it decrease} of PCR (with increasing the magnetic
field) in some range of the currents for photons with wavelengths
$\lambda$ =450-1200 nm. Found experimental results for all studied
wavelengths $\lambda = 450-1550$ nm could be explained by the
vortex hot spot model (which assumes partial suppression of
$\Delta$ in the area with size smaller than the width of the
nanowire) if one takes into account nucleation and entrance of the
vortices to the photon induced hot spot and their pinning by the
hot spot with relatively large size and strongly suppressed
$\Delta$.

\end{abstract}

\maketitle

\section{Introduction}

The main idea of superconducting nanowire single photon detector
(SNSPD) is based on destruction of the superconductivity by the
absorbed high-energy photon (which produces hot quasiparticles) in
relatively large area of the superconducting nanowire
\cite{Semenov_first}. The appearance of such a region decreases
the superconducting properties of the nanowire and leads to the
resistive state (if transport current $I$ exceeds some critical
value) which is visible via appearance of the voltage. First
realization of such a detector was done in 2001 \cite{SSPD_first}
and since that time there were many theoretical
\cite{Semenov_hot_spot,Maingault,Zotova_PRB,Bulaevskii,Eftekharian,Engel_JAP,Zotova_SUST}
and experimental works (for review see \cite{Natarajan}) which
were aimed to understand the physical details of their working
mechanism and to improve their characteristics.

Despite clear main idea why SNSPD works there are still debates
about details of the appearance of the resistive state in SNSPD.
These debates are connected with absence of rigorous study (which
needs solution of kinetic equations coupled with equation for
superconducting order parameter) of initial stage of the response
of the superconducting nanowire on the absorbed photon. In the set
of theoretical works
\cite{Semenov_first,Semenov_hot_spot,Eftekharian,Engel_JAP}
authors used approach, which is similar to Rothwarf-Taylor model
\cite{RT} and those quantitative validity is questionable in case
when there is spatial gradient of superconducting order parameter
$\Delta$. Nevertheless in the literature one may find two ideas
about what happens after photon absorption: the photon creates hot
spot or hot belt. According to the hot spot model absorbed photon
creates region with locally suppressed superconductivity
(partially or completely) with the size smaller than the width of
the nanowire
\cite{Semenov_hot_spot,Maingault,Zotova_PRB,Eftekharian,Engel_JAP,Zotova_SUST}.
As a result the superconducting current has to crowd near the hot
spot and in Refs. \cite{Zotova_PRB,Engel_JAP,Zotova_SUST} it is
argued that the superconducting state becomes unstable at the
current $I>I_{det}$ (which is smaller than the critical current of
the nanowire without hot spot) due to nucleation and motion of the
vortices.

Authors of the hot belt model \cite{Bulaevskii} assume that the
absorbed photon creates the spatially uniform distribution of hot
quasiparticles across whole width of the nanowire and it results
in the appearance of some kind of weak link. The critical current
$I_c^{belt}$ of the nanowire with hot belt (weak link) is smaller
than the critical current of the nanowire $I_c$ and when
$I_c^{belt}<I<I_c$ the resistive state is realized after
absorption of the photon. Vortices are involved in the hot belt
model in order to explain smooth dependence of the detection
efficiency (DE) of SNSPD on the applied current (seen in all
experiments on SNSPD) -- the finite DE at $I<I_c^{belt}$ in this
model is connected with thermoactivated vortex entrance via edges
of the nanowire and their motion heats the nanowire and provides
large voltage response. Note that in the hot spot model considered
in Ref. \cite{Zotova_SUST} the smooth dependence $DE(I)$ follows
not only from the thermoactivated vortex entrance to the hot spot
but also from dependence of $I_{det}$ on the transverse coordinate
$x$ of hot spot in the nanowire.

In further we use term intrinsic detection efficiency (IDE), which
describes probability to have the resistive response in the
superconducting nanowire after photon absorption (IDE$=1$ means
that each absorbed photon produces resistive response). Detection
efficiency in real detectors is always smaller than IDE because
absorption in the detector is less than unity. With this
definition IDE is the intrinsic characteristic of the
superconducting nanowire and not the whole detection system.
Experimentally dependence IDE(I) could be found if the photon
count rate (PCR) saturates at large currents and
IDE(I)$=$PCR(I)/PCR$_{sat}$.

To distinguish experimentally which of these two models is more
relevant to the detection mechanism of SNSPD one may use external
magnetic field. Hot belt model predicts parallel shift of
dependence PCR(I) (or IDE(I)) in direction of small currents with
increasing magnetic field. This result directly comes from decay
of $I_c^{belt}$ in the magnetic field -- well know result
following from the theory of Josephson junctions \cite{Tinkham}
and narrow superconducting films \cite{Stejic}. The hot spot model
of Ref. \cite{Zotova_SUST} predicts more complicated behavior,
with noticeable shift of IDE(I) at low currents (where finite $IDE
\ll 1$ is assumed to be finite due to thermoactivated vortex
entrance to the hot spot - like in the hot belt model) and much
weaker field dependence at the currents where $IDE \gtrsim 0.1$
and vortices appears in the nanowire at the current $I>I_{det}(x)$
without any fluctuations. Moreover this model predicts {\it
decrease} of PCR at the currents, where PCR saturates.

Recent experiment on MoSi based SNSPD \cite{Korneev_MoSi} discards
the hot belt model for this detector. It was found that for high
energy photon ($\lambda =450$ nm) detection efficiency practically
did not vary in the wide range of the magnetic fields both at the
currents where IDE $\ll 1$ and IDE $\sim 1$. This result also
discards the possibility of photon detection via thermoactivated
vortex entry in the hot spot model of Ref. \cite{Zotova_SUST}. For
photons with larger wavelength ($\lambda$=600--1000 nm) in Ref.
\cite{Korneev_MoSi} dependence $DE(I)$ shifted in direction of
smaller currents with increase of the magnetic field. The found
result -- the larger $\lambda$ the larger shift is also in
contrast with prediction of the hot belt model where one could
expect larger dependence on magnetic field for photons with larger
energy (in this case $\Delta$ should be suppressed stronger in the
hot belt region and one needs smaller magnetic field to suppress
$I_c^{belt}$). In Ref. \cite{Zotova_SUST} no study was made for
field-dependent IDE for photons with different energies.

In the present work we study effect of the magnetic field on
dependence PCR(I) of SNSPD based on NbN. Among different studied
detectors there was one which had saturation of PCR(I) in large
range of the wavelengths $\lambda$ = 450--1000 nm. Note, that for
MoSi based SNSPD in Ref. \cite{Korneev_MoSi} the saturation of
PCR(I) was observed only for photons with $\lambda=450$ nm but it
was not fully complete (PCR still slightly increased with current
increase). Qualitatively for all studied NbN detectors we find the
same results as for the MoSi based SNSPD. Besides due to high
quality of one of the studied detectors (mentioned above) we find
effect which was not observed for MoSi SNSPD and which was
theoretically predicted in Ref. \cite{Zotova_SUST} -- external
magnetic field decreases PCR (when it is close to PCR$_{sat}$ at
zero magnetic field) and shifts the current where PCR saturates to
larger values. To explain different field dependence of PCR(I)
(IDE(I)) observed for different wavelengths both for MoSi and NbN
detectors we modified the hot spot model of Ref.
\cite{Zotova_SUST} and assumed that the resistive state starts
only if the vortex, which enters the hot spot via edge of the
nanowire, can freely pass the hot spot region (i.e. it should be
unpinned). Using this assumption and the model of Ref.
\cite{Zotova_SUST} we were able to explain observed field
dependence of PCR(I) at all studied wavelengths for both
materials.

\section{Experiment}

\begin{widetext}
\begin{table}
\caption{\label{tab:table1}Material and physical properties (at
$T=1.7K$) of studied NbN based SNSPD.}
\begin{ruledtabular}
\begin{tabular}{cccccccccc}
  Sample & w(nm)& d(nm)& T$_c$(K) & I$_c(0)(\mu A)$ & R$_{sq}$(Ohm) & $\xi$(nm) & I$_{dep}^{th} (\mu A)$ & $\beta_{exp} $ & $\beta_{th} $ \\
NbN1& 102 & 4 & 10.1 & 27.1 & 512 & 4.9 & 44 & 108 & 208 \\
NbN2& 90  & 6 & 9.9  & 21.2 & 456 & 4.7 & 45 & 85  & 227 \\
NbN3& 110 & 5 & 9.2  & 25.1 & 420 & 4.7 & 57 & 83  & 232 \\
\end{tabular}
\end{ruledtabular}
\end{table}
\end{widetext}

In our experiments we used a cryoinsert with superconducting
solenoid for the storage dewar in which operation temperature 1.7
K was achieved by helium vapor evacuation. The magnetic fields
from 0 to 425 mT were applied perpendicular to the SNSPD plane.
The SNSPD was fixed to a sample holder in a dipstick and was
wire-bounded to a transmission line with a coplanar-coaxial
connector. The SNSPD-chip with the transmission line was connected
to DC+RF-output port of a bias-T. DC port was connected with
precision voltage source. Absorption of a photon produces a
voltage pulse which is amplified by two room-temperature
amplifiers Mini-Circuits ZFL-1000LN+ (1 GHz band, 46 dB total
amplification) and it is fed to a digital oscilloscope and a pulse
counter. We recorded the count rate during an 1 s interval at each
current. As a light source we used a grating spectrometer with a
black body for wavelength from 400 to 1550 nm. The light is
delivered to the SNSPD by a optical fiber SMF-28e with 9 $\mu$m mode
field diameter. The meanders were precisely aligned against the
fiber core and illuminated from the top side.

We studied three NbN detectors with different material and
physical parameters (see the table 1). Only one of the detectors
(NbN1) shows well determined saturation of the photon count rate
at large currents in wide range of wavelengths $\lambda=450 -1000$
nm (see Fig. 1) and we mainly present results for this detector.
In Fig. 1 we use linear scale to demonstrate the clear saturation
of PCR and maximal IDE $\simeq 1$. Only for wavelengths
$\lambda=1200$ and $1550$ nm intrinsic detection efficiency
IDE=PCR(I)/PCR$_{sat}$ does not reach unity (for these wavelengths
we find PCR$_{sat}$ by extrapolation of the experimental data at
larger currents using similarity between shapes of PCR(I) for
different wavelengths - see Fig. 1).
\begin{figure}[hbtp]
\includegraphics[width=0.53\textwidth]{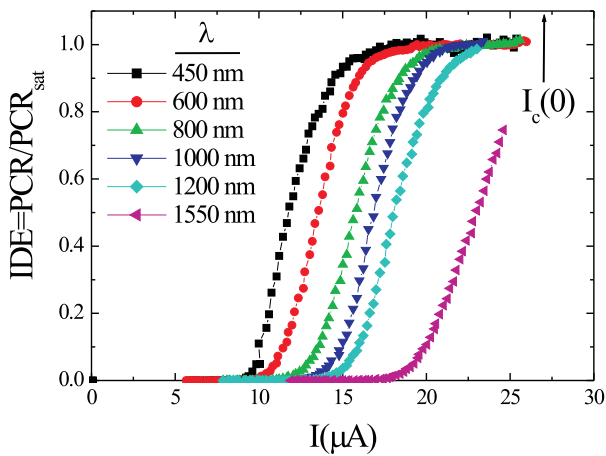}
\caption{Current dependence of the normalized photon count rate
(PCR/PSR$_{sat}$(H=0)=IDE) for different wavelengths found for
good quality detector NbN1. For $\lambda$=1200 and 1550 nm we find
PSR$_{sat}$(H=0) from extrapolation of the experimental results to
larger currents and assuming small variation of the shape of
dependence IDE(I) with change of $\lambda$.}
\end{figure}

In Fig. 2 we show effect of the external magnetic field $H$ on the
photon count rate for photons with different energy (wavelength).
Also as for Fig. 1 we subtract dark counts those rate rapidly
grows when current approaches the critical current of the meander
$I_c(H)$ (its dependence on $H$ is shown in the inset in Fig.
2(d)). Effect of magnetic field is very similar to the one found
before in Ref. \cite{Korneev_MoSi} for MoSi detector -- with
increase of $H$ dependence PCR(I) shifts to the direction of {\it
small}  currents (at fixed current it means increase of PCR), and
this shift is smaller the higher the energy of the photon. The new
effect which was not observed before is the {\it decrease} of PCR
(when IDE $\gtrsim 0.5$) at currents larger than some 'crossover'
current $I_{cross}$  - see Fig. 2. At large magnetic fields the
effect is not visible because $I_c(H)$ rapidly decreases and dark
counts interfere the photon counts before IDE reaches $\simeq
0.5$.

We observed the same crossover behavior also for other NbN
detectors for photons with $\lambda$=500-800 nm when PCR saturated
at $H=0$. For MoSi detector the crossover was not found and
dependence IDE(I) just shifted to the direction of small currents
(for this specific detector PCR did not saturate even for photons
with $\lambda=450 nm$).
\begin{figure}[hbtp]
\includegraphics[width=0.53\textwidth]{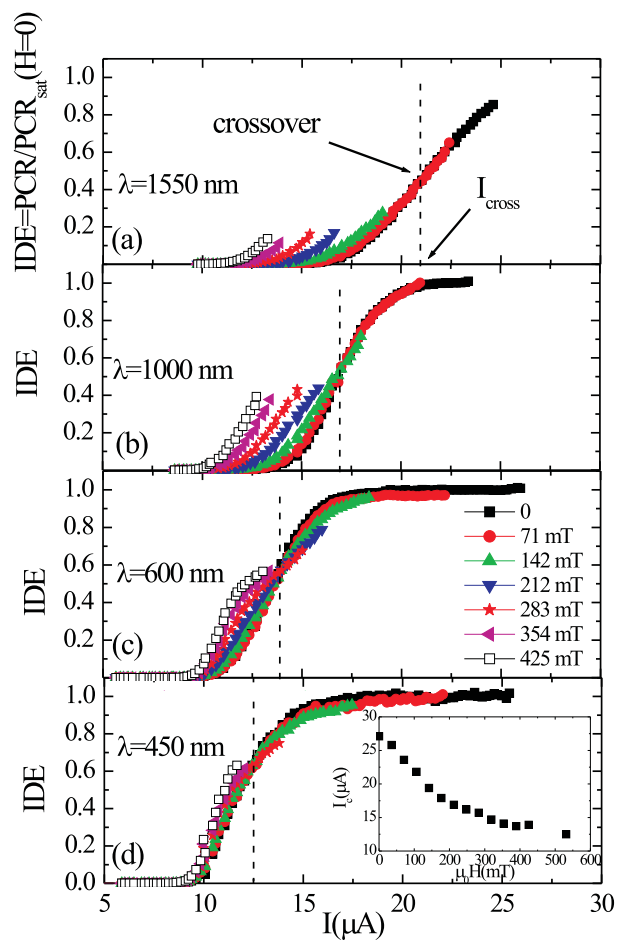}
\caption{IDE(I) of NbN1 detector at different magnetic fields. In
the inset in figure (d) we present field dependence of the
critical current of the meander.}
\end{figure}

Dependence of dark count rate (DCR) on the magnetic field (see
Fig. 3) is in strict contrast with the results shown in Fig. 2.
DCR follows the change in the critical current of the meander and
curve DCR(I) shifts to smaller currents with increasing $H$.
Notice, that the shift is well visible even for $\mu_0H \leq 71$
mT when photon count rate practically does not change for all
studied wavelengths. On the contrary, at large magnetic fields DCR
slightly depends on H (because of small change in the critical
current -- see inset in Fig. 2(d)), while photon count rate shows
noticeable field dependence (at least for photons with $\lambda=$
1000 -- 1500 nm). The present results are very similar to the
results found for other NbN and MoSi detectors (see Fig. 4 below
and Fig. 4 in Ref. \cite{Korneev_MoSi}). They demonstrate that the
dark counts are most probably originated from the thermo-activated
vortex entrance near the 'weakest' place of the meander which
determines its critical current. We may conclude it from very fast
decay of DCR with the current which is consequence of large
increase of the energy barrier for vortex entry $\delta F$ as
current decreases
\cite{Qiu,Bartolf,Bulaevskii_barrier,Vodolazov_barrier}. For all
studied detectors the current dependence of dark count rate could
be fitted by linear function $ln(DCR)=-\beta_{exp}(1-I/I_c(H))$
where the coefficient $\beta_{exp}$ found at $H=0$ is shown in
table 1.
\begin{figure}[hbtp]
\includegraphics[width=0.48\textwidth]{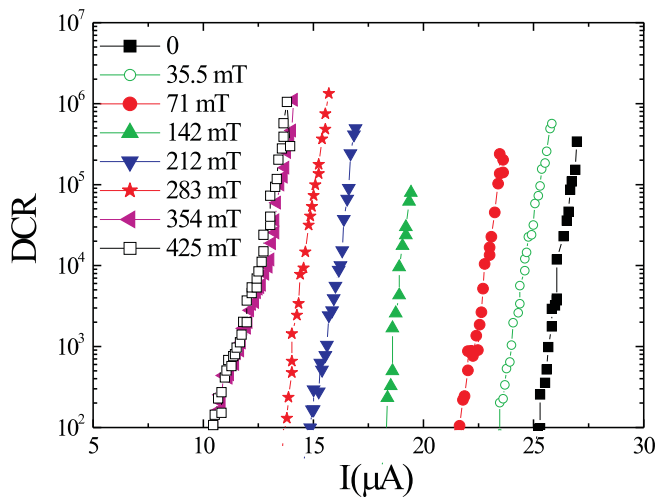}
\caption{Dark count rate in NbN1 detector at different magnetic
fields.}
\end{figure}

Before comparison of $\beta_{exp}$ with predictions of theoretical
models we have to stress that the 'weakest' place could be located
not only at the edges but also in the middle of straight pieces of
the meander. In Fig. 4 we show dark count rate and critical
current of NbN2 detector measured at different magnetic fields.
One can see that at low magnetic fields ($\mu_0 H\leq $36 mT) the
critical current practically does not depend on $H$ and DCR does
so. Plateau in dependence $I_c$(H) as $H \to 0$ may appear if
there is relatively large intrinsic defect in the middle of the
nanowire and resistive state starts via nucleation of the
vortex-antivortex pair in that place (see discussion in section
IIIB near Fig. 9). In this situation the smallest energy barrier
corresponds to the nucleation of the pair vortex-antivortex near
the intrinsic defect and not to the single vortex entry via edge
of the nanowire.

Therefore to make quantitative comparison one needs to know exact
parameters of the 'weakest' place and where it is located in the
meander. Because we do not have such an information we make rough
estimation using the result following from the London model for
the single vortex entry to the straight nanowire {\it without edge
or bulk defects} $\beta_{th}=\Phi_0^2d/16\pi^2 \lambda_L^2 k_BT$
(see Eq. (2) in Ref. \cite{Vodolazov_barrier}, where $\lambda_L$
is the London penetration depth). Calculated $\beta_{th}$ are
shown in the table 1. One can see the large quantitative
difference between $\beta_{exp}$ and $\beta_{th}$ which we connect
with effect of unknown parameters of the 'weakest' place on value
of the energy barrier.
\begin{figure}[hbtp]
\includegraphics[width=0.48\textwidth]{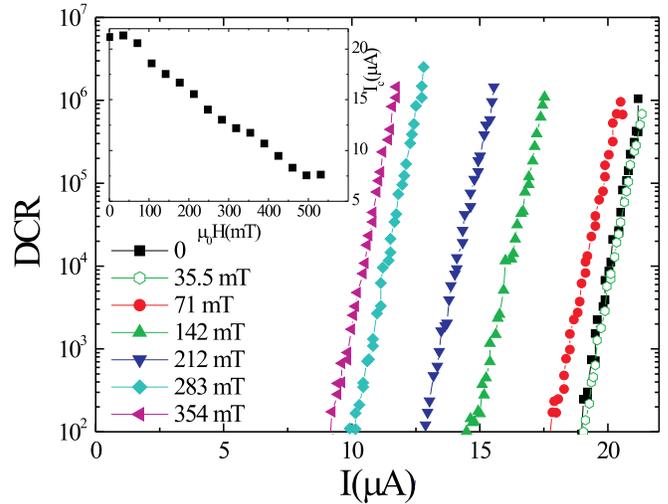}
\caption{Dark count rate and critical current (in the inset) of
NbN2 detector at different magnetic fields.}
\end{figure}

\section{Theory}

To calculate dependence IDE(I,H) we use the model of Ref.
\cite{Zotova_SUST}. The hot spot (HS) is modelled as a region in
the form of a circle with radius $R$ and inside this area the
quasiparticle distribution function $f(\epsilon)$ deviates from
the equilibrium (the quasiparticles are 'heated')
\cite{Zotova_SUST}. Because of 'heated' quasiparticles the
superconducting order parameter $\Delta=|\Delta|e^{i\varphi}$ is
suppressed inside the hot spot and the critical current of the
nanowire changes. Our aim is to find its value (we call it
detection current $I_{det}$ to distinguish it from the critical
current of nanowire without hot spot) by solving the
Ginzburg-Landau equation for $\Delta$
\begin{equation}
\xi_{GL}^2\left(\nabla-\frac{2ieA}{\hbar
c}\right)^2\Delta+\left(1-\frac{T}{T_c}+\Phi_1-\frac{|\Delta|^2}{\Delta_{GL}^2}\right)\Delta=0.
\end{equation}
The term \cite{Ivlev,Larkin,Kramer}
\begin{equation}
\Phi_1=\int_{|\Delta|}^{\infty}\frac{2(f^0-f)}{\sqrt{\epsilon^2-|\Delta|^2}}d\epsilon
,
\end{equation}
describes the effect of nonequilibrium distribution function
$f(\epsilon)\neq f^0(\epsilon)=1/(exp(\epsilon/k_BT)+1)$ on
$\Delta$. In Eq. (1) $\xi_{GL}^2=\pi\hbar D/8k_BT_c$ and
$\Delta_{GL}^2=8\pi^2(k_BT_c)^2/7\zeta(3)\simeq 9.36(k_BT_c)^2$
are zero temperature Ginzburg-Landau coherence length and
superconducting order parameter correspondingly. For numerical
calculations it is convenient to write Eq. (1) in dimensionless
units (length is scaled in units of
$\xi(T)=\xi_{GL}/(1-T/T_c)^{1/2}$, $\Delta$ in units of
$\Delta_{eq}=\Delta_{GL}(1-T/T_c)^{1/2}$ and vector potential $A$
in units of $\xi H_{c2}$, where $H_{c2}$ is the second critical
magnetic field)
\begin{equation}
(\nabla-i\tilde{A})^2\tilde{\Delta}+\left(\alpha-|\tilde{\Delta}|^2\right)\tilde{\Delta}=0,
\end{equation}
with $\alpha=(1-T/T_c+\Phi_1)/(1-T/T_c)$.

In our model we have two control parameters: radius of the hot
spot and value of $\Delta$ inside HS ($\Delta_{in}$), which is
controlled by the parameter $\alpha$ in Eq. (3) (or $\Phi_1$ in
Eq. (1)). In contrast with other hot spot models (see for example
Refs.
\cite{Semenov_first,Semenov_hot_spot,Bulaevskii,Eftekharian,Engel_JAP})
our approach automatically resolves question about stability of
the superconducting state of the nanowire without usage of extra
assumptions (like additional condition for vortex entry needed in
the London model) and it takes into account the current continuity
equation $div j=0$ (which comes from the imaginary part of Eq. (1)
or Eq. (3)). Drawback of our approach is the unknown quantitative
relation between the energy of the absorbed photon and the size of
the hot spot and how strong $\Delta$ is suppressed inside it
(strictly speaking is has to be found from the solution of the
kinetic equation for $f(\epsilon)$ coupled with the equation for
$\Delta$). Despite it our theoretical results explain the
experimental field dependence of the photon count rate (see
results below) and it gives us the hope that the used model
captures the essential physics of the detection mechanism of
single photons in SNSPD.
\begin{figure}[hbtp]
\includegraphics[width=0.53\textwidth]{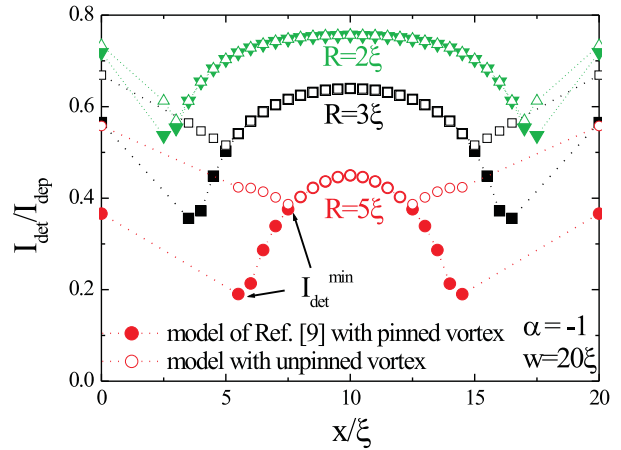}
\caption{Dependence of the detection current $I_{det}$ (at which
the resistive response appears in the SNSPD after photon
absorption) on the coordinate of the hot spot, defined in two
models: model with pinned vortex \cite{Zotova_SUST} (solid
symbols) and model with unpinned vortex (empty symbols).}
\end{figure}

\subsection{Straight nanowire}

In simulations we place the hot spot in different places across
the straight nanowire and find dependence $I_{det}(x)$ via
numerical solution of Eq. (3) (details of the numerical scheme are
present in Ref. \cite{Zotova_SUST}). When the photon is absorbed
at the edge of the nanowire we model the hot spot by a semicircle
with radius in $\sqrt{2}$ times larger than radius of the hot spot
inside the nanowire \cite{Zotova_PRB,Zotova_SUST}. The resistive
state is realized via nucleation of the vortices (they enter via
edge of the nanowire when HS is located close to the edge and
vortex-antivortex pair is nucleated inside HS when it is located
near a center of the nanowire) and their motion across the
superconductor \cite{Zotova_SUST}. In Ref. \cite{Zotova_SUST} it
was found that relatively large hot spot can pin the vortex when
it enters the nanowire and one needs to enlarge current to make it
unpinned. Unlike usual pinning center the hot spot relaxes in time
and if it contains the vortex then at some moment in time the
vortex could be unpinned and moves across the nanowire and heats
it. Based on this idea the detection current in Ref.
\cite{Zotova_SUST} was defined as a current at which the vortex
enters the hot spot when it is located close to the edge of the
nanowire.
\begin{figure}[hbtp]
\includegraphics[width=0.46\textwidth]{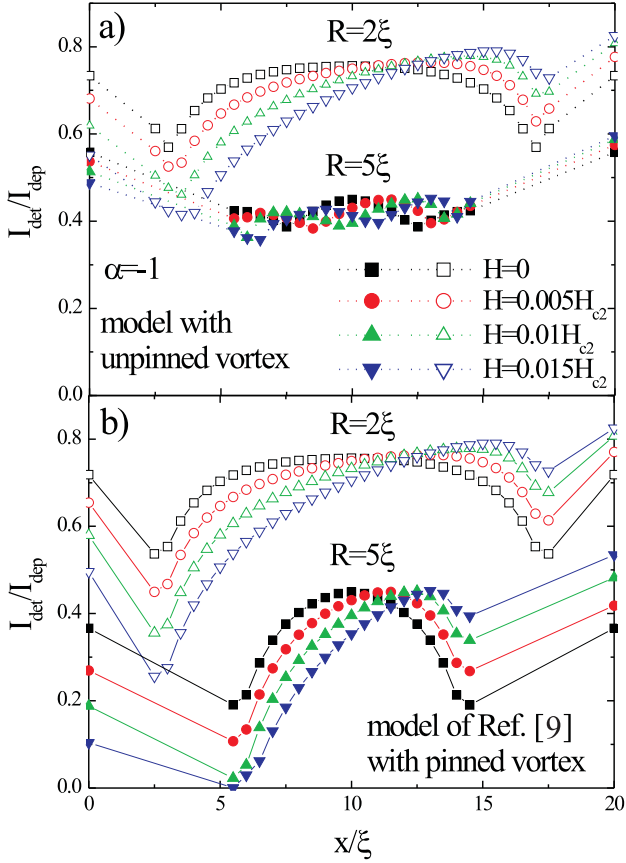}
\caption{Dependence of the detection current $I_{det}$ on the
coordinate of hot spot with two radii at different magnetic
fields. Figure a) corresponds to the model with unpinned vortex
and b) - to the model with pinned vortex  of Ref.
\cite{Zotova_SUST}.}
\end{figure}

In the present work we define $I_{det}$ as the current at which
the permanent motion of the vortices starts in the nanowire (i.e.
vortex has to be unpinned from the hot spot). To see the
difference of present definition of $I_{det}$ with one proposed in
Ref. \cite{Zotova_SUST}, in Fig. 5 we show dependence of both
$I_{det}$ on the coordinate of the hot spot. One can see that the
difference exists when the hot spot is located near the edges of
the nanowire and with increase of radius of HS the difference
increases. It occurs due to enhanced pinning ability of the hot
spot, which in general depends also on $\Delta_{in}$ (it is
controlled by the parameter $\alpha$ in our model). For example
when $\alpha=0$ the hot spot with $R=2\xi$ already cannot pin
vortices (at least at $H=0$ and width of the nanowire $w=20\xi$)
and the current, at which vortex enters the hot spot coincides
with the current when it becomes unpinned. For $R=5\xi$ the hot
spot cannot pin vortex nowhere in the nanowire when $\alpha \geq
0.36$ (it corresponds to $\Delta_{in}\geq 0.6 \Delta_{eq}$ inside
HS).
\begin{figure}[hbtp]
\includegraphics[width=0.53\textwidth]{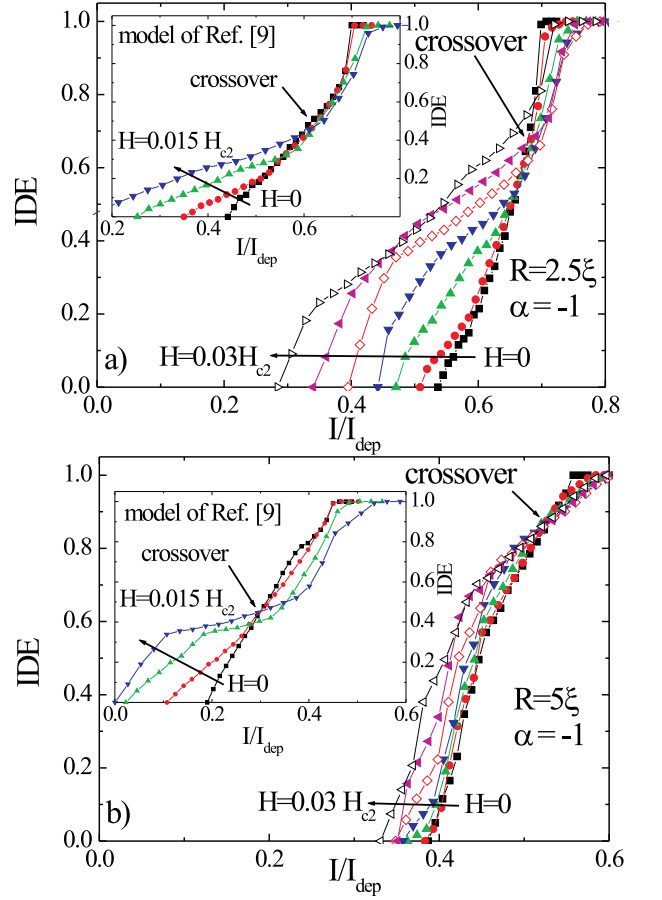}
\caption{Current dependence of IDE at different magnetic fields
(from $H=0$ up to $H=0.03H_{c2}$ with step $\delta H=0.005
H_{c2}$) following from the vortex hot spot model with unpinned
vortex. Figure (a) corresponds to the hot spot with radius $R=2.5
\xi$ while figure (b) to the hot spot with $R=5\xi$ (in both cases
$\alpha=-1$). In the insets similar dependencies are calculated in
the hot spot model with pinned vortex of Ref. \cite{Zotova_SUST}.}
\end{figure}

In Fig. 6 we show how $I_{det}$ changes in the magnetic field. In
the model with unpinned vortex the minimal detection current
$I_{det}^{min}$ changes slightly in weak magnetic fields
($H<H_s\simeq \Phi_0/2\pi \xi w$, with $H_s\simeq 0.05 H_{c2}$ for
the nanowire with $w=20 \xi$) when the radius of the hot spot is
large and vortex pinning is strong. Physical reason for the found
effect is following - external magnetic field favors the vortex
entry to HS because of increase of the current density at one edge
of the nanowire \cite{Zotova_SUST} but it weakly changes the
current density near nanowire's center which is important from
point of view of vortex unpinning when the hot spot is located
near the same edge. When the pinning ability by HS is weak (as for
the hot spot with radius $R=2\xi$) the change of $I_{det}^{min}$
is large and it practically follows the change of the current
density at the edge of the nanowire due to external magnetic
field.

The intrinsic detection efficiency at given current $I$ in our
model is determined as a photon-sensitive part of the nanowire
where $I>I_{det}(x)$ \cite{Zotova_SUST}. In Fig. 7 we present
calculated IDE at different magnetic fields (magnetic field $0.005
H_{c2}$ for NbN1 detector with $\xi(T=1.7K)=4.9 nm$ corresponds to
$\simeq 69 mT$) and different radii of hot spot.

First of all from Fig. 7 one can see that in both models and for
all radii of the hot spot there is 'crossover' current above which
IDE {\it decreases} while at $I<I_{cross}$ IDE {\it increases}
with increase of the magnetic field. The first effect comes from
the increase of maximal detection current $I_{det}^{max}$ while
the second effect originates from decrease of $I_{det}^{min}$ (see
Fig. 6). Secondly, in the model with pinned vortex there is strong
dependence of IDE on the magnetic field (see insets in Fig. 7) for
all studied radii of HS while in the model with unpinned vortex
the field dependence is determined by its radius - the larger the
radius the weaker field dependence.

In addition we considered hot spots with {\it fixed} radius $R=5
\xi$ and varying coefficient $\alpha$ in range $0-0.64$, which
corresponds to $\Delta_{in}$=$0-0.8 \Delta_{eq}$ (in absence of
transport current). We find that for all values of $\alpha$ there
is 'crossover' current and with decreasing $\alpha$ field
dependence of IDE(I) becomes weaker (again due to increased
ability of the hot spot to pin the vortices).

We also studied how presence of edge defects may influence photon
count rate and IDE. First of all we considered uniform suppression
of the critical temperature along nanowire's edges (in the area
with width $\delta x=\xi/2$) where we put $\alpha=-1$. Such a
'dead' layer influences quantitatively the shape of dependence
IDE(I,H) (compare Fig. 7 and Fig. 8). We also calculate IDE(I,H)
for another type of the edge defect (point like suppression of
$T_c$ at the edge) and different values of coefficient $\alpha$
inside the hot spot. All these changes leaded to quantitative
variation of IDE(I,H) (not shown here) but qualitative properties
(namely the presence of the 'crossover' current and dependence of
IDE(H) on the radius of the hot spot and/or $\Delta_{in}$) stay
the same.
\begin{figure}[hbtp]
\includegraphics[width=0.46\textwidth]{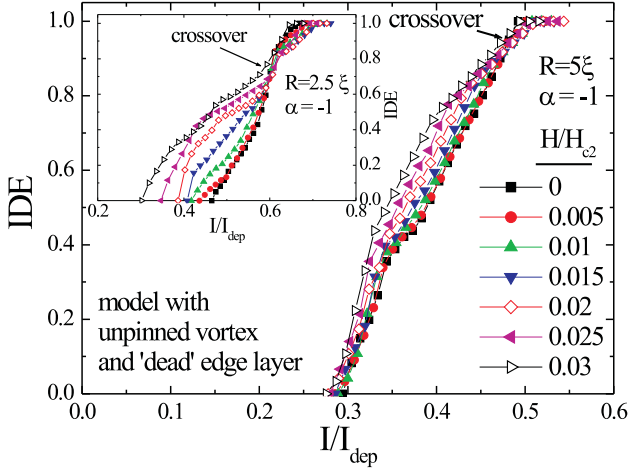}
\caption{Current dependence of IDE at different magnetic fields
following from the vortex hot spot model with unpinned vortex and
'dead' edge layers in the nanowire, where $T_c$ is locally
suppressed.}
\end{figure}

\subsection{Effect of meander geometry and bulk intrinsic defects}

We have to stress, that the results present in Figs. 5-8
correspond to the straight nanowire, while real detectors have a
form of the meander. In Fig. 9 we show calculated field dependence
of the critical current of the meander with parameters close to
experimental ones (see insets in Fig. 9) and straight nanowire
with width $w=20\xi$. We choose such a bend of the meander where
the magnetic field suppresses the critical current (for the
opposite bend or the opposite current direction the same magnetic
field may enhance $I_c$ \cite{Clem_corner}). Because of the
current crowding effect $I_c(H)$ of the meander is substantially
smaller than $I_c(H)$ of the straight nanowire \cite{Clem_corner}
(the difference is not large at zero magnetic field but it is more
pronounced at finite $H$) and it makes unreachable large values of
IDE calculated at finite $H$ and shown in Figs. 7-8 for straight
nanowire. When current approaches $I_c(H)$ the dark counts will
interfere the photon counts and SNSPD stops to work.
\begin{figure}[hbtp]
\includegraphics[width=0.48\textwidth]{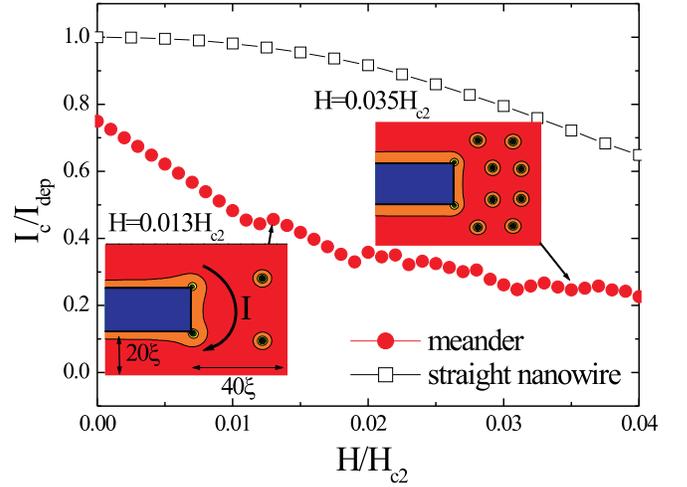}
\caption{Magnetic field dependence of the critical current of the
straight nanowire with w=20 $\xi$ (empty squares) and the meander
(solid circles) with geometrical parameters shown on the bottom
inset. For the meander we take into account the bend where
external magnetic field decreases the critical current. In the
insets we show contour plot of $|\Delta|$ for the meander at
different magnetic fields and $I\lesssim I_c(H)$. At $H \leq 0.012
H_{c2}$ there are no vortices in the meander. In the narrowest
place of the meander vortices appear at $H>0.038 H_{c2}$.}
\end{figure}

It is interesting to note, that according to our calculations
there are no vortices in the narrowest part of the meander at
$I\lesssim I_c(H)$ up to the field $H=0.038 H_{c2}$. On the
contrary vortices enter the bend region at field $H \geq 0.013
H_{c2}$. It occurs due to wider part of the meander in the bend
region and smaller value of the critical magnetic field $H_s
\simeq \Phi_0/\xi w$ at which vortices may enter superconducting
nanowire \cite{Stejic}.

As we show in the inset of Fig. 4 dependence $I_c(H)$ of real
meander at $H\to 0$ may have a plateau instead of linear decay.
Such a dependence is possible if somewhere in the middle of
straight pieces of the meander there is relatively large intrinsic
defect (where locally either $T_c$ is suppressed or nanowire is
thinner) \cite{self}. From physical point of view the effect of
intrinsic defect on $I_c$ is similar to the effect of hot spot on
$I_{det}$. From Fig. 6 it follows that if intrinsic defect (or hot
spot) is located close to the center of straight nanowire it
provides much weaker field dependence of $I_c$ than the similar
defect placed at the edge. Only when the current density
(supervelocity) at the nanowire's edge exceeds (at some magnetic
field) the current density (supervelocity) near the central defect
then vortices start to enter the nanowire via edge (and not the
vortex-antivortex pair is nucleated at the central defect) and
critical current recovers strong field dependence.

Intrinsic defects may influence $I_c$ not only at low magnetic
fields but also at relatively large fields, when vortices exist in
the meander at $I \lesssim I_c(H)$. Indeed, calculated $I_c(H)$
decays faster than the experimental one (compare Fig. 9 with inset
in Fig. 2(d)) at the magnetic fields $H \gtrsim 0.015 H_{c2}$
($\mu_0H \gtrsim 207 mT$ for NbN1 sample) when $I_c$ changes
nonlinearly with $H$. Such a deviation could be explained by the
vortex pinning at the intrinsic defects of the nanowire which are
absent in our calculations. Similar effect was observed for NbN
straight films in the recent experiment \cite{Ilin} and
analytically it was calculated in Ref. \cite{Elistratov} for the
bulk pinning described by the Bean model. Vortex pinning should
prevent fast decay of $I_{det}^{min}$ with increase of $H$ (in
this case the vortex should overcome not only the pinning by the
hot spot but also the intrinsic pinning outside HS) and it has to
be taken into account for quantitative comparison of experimental
and theoretical IDE(I) at magnetic fields larger than $0.015
H_{c2}$ ($\sim 207 mT$) for detector NbN1.

Now we would like to discuss contribution of the bends to
detection ability of the meander-like detectors. Because of the
current crowding effect the regions near the interior corners of
the bends are capable to detect photons at the currents lower than
the minimal detection current $I_{det}^{min}$ of the straight
nanowire. Rough estimation, based on the difference between
critical current of the meander and the straight nanowire says
that near-corner area stops to detect photons at the current
$I^{min}=I_{det}^{min} \cdot I_c^{meander}/I_c^{nanowire}$. This
estimation works relatively well for photons which create small
hot spots (with radius $R\lesssim 2\xi$) while for larger spots
with small $\Delta_{in}$ it underestimates $I^{min}$ and $I^{min}$
lies closer to $I_{det}^{min}$ (numerical calculations in Ref.
\cite{Zotova_SUST} confirm it - see Fig. 12 there).

Using numerical simulations we find in what area S near the bend
the local current density is larger (we use criteria
$j>1.01j_{\infty}$) than the current density far from the bend
$j_{\infty}$. For our geometrical parameters (see Fig. 9) we find
$S \simeq 400 \xi^2 \sim w^2$. For parameters of NbN1 meander it
consists about $1.4\%$ of all area of the detector and it gives us
the rough estimation for the photon-sensitive area at the currents
just below $I_{det}^{min}$. When current decreases further this
area shrinks as $I\to I^{min}$. Therefore we suggest that finite
$0<$IDE $\lesssim$ 0.014 in the current range
$I^{min}<I<I_{det}^{min}$ comes from this photon-sensitive area
located near the interior corners of the meander. This suggestion
is also supported by very weak field dependence of IDE(H)$\ll 1$
observed in our experiment at $\mu_0 H<70 mT$. If finite and small
IDE would be connected with thermoactivated vortex entry (as it
was assumed in \cite{Bulaevskii,Zotova_SUST}) it should have
strong field dependence in the same range of magnetic fields as a
dark count rate has (see Figs. 3,4). But if our idea is correct,
then IDE should not change (or change very slightly) because of
presence of the {\it right} and {\it left} bends in the meander.
Indeed, external magnetic field increases the current density in
one kind of bend (let it be right one for definiteness) and
decreases it in another one. We calculate the change in the area
near both bends where current density is locally enhanced and find
that this area practically does not vary at low magnetic fields
because its expansion near the right bend is compensated by its
shrinkage near the left bend.

\subsection{Threshold current versus photon's energy}

In Fig. 10 we show experimental dependence of the current (we call
it as threshold current $I_{thr}$), at which IDE reaches 0.9, on
the energy of the photon $E$. Our choice of the cut-off is not
accident. Indeed, in the bend region there is large area, where
the current density is smaller than $j_{\infty}$ and this part
participates in photon detection at larger currents than the
straight pieces of the meander. Roughly, for our geometrical
parameters this area is about $10 \%$ of area of the meander.
Hence, when the current approaches $I_{det}^{max}$ of the straight
nanowire the intrinsic detection efficiency of the meander reaches
0.9 and to reach IDE$=1$ one should increase current further.
\begin{figure}[hbtp]
\includegraphics[width=0.46\textwidth]{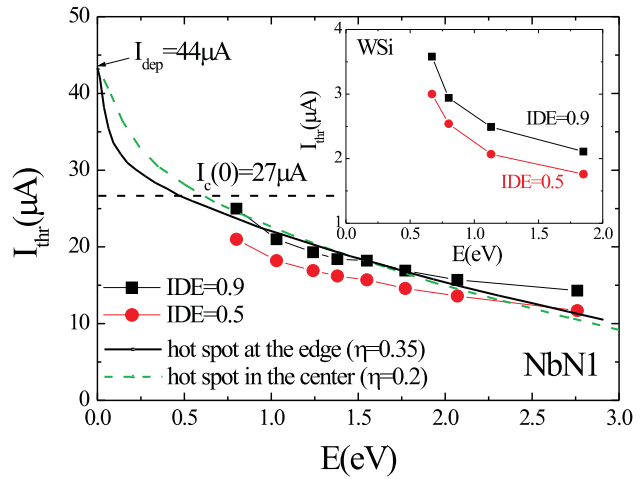}
\caption{Dependence of the current, at which IDE reaches 0.9
(squares) and 0.5 (circles), on the energy of the photon (results
are obtained for NbN1 detector). The theoretical curves (solid and
dashed) are found in assumption that current $I_{thr}$ corresponds
to the hot spot located in the center of the nanowire or at its
edge (in the last case it has form of semicircle). Fitting
coefficient $\eta$ describes what part of the photon's energy goes
for suppression of $\Delta$ inside the hot spot
\cite{Vodolazov_PRB}. In the inset we show results for WSi based
detector extracted from Fig. 2 of Ref. \cite{Baek}.}
\end{figure}

In Fig. 10 we also plot theoretical dependencies, following from
the vortex hot spot model developed here. We use two locations of
the hot spot (in the center and at the edge of the nanowire) and
fitting parameter $\eta$, which describes what part of the energy
of the photon goes for suppression of $\Delta$ inside the hot spot
\cite{Vodolazov_PRB}. The quantitative agreement between the
theory and the experiment is poor, which justifies that the used
model assumptions are too rough. Indeed, the shape of the hot spot
is not obligatory should has a round shape, because in the
presence of the transport current it will preferably grows in the
direction perpendicular to the current flow (due to current
crowding at the equator of HS). Besides the coefficient $\eta$ may
depend on the energy of the photon. Both these factors are not
taken in our model, because they need calculation of the dynamics
of nonequilibrium quasiaprticles and solution of the kinetic
equation. However our model predicts nonlinear dependence of
$I_{thr}(E)$ with rapid grow of $I_{thr}$ up to $I_{dep}$ as $E
\to 0$ which resembles experimental results.

Our experimental dependence $I_{thr}(E)$ drastically differs from
the linear relation found in Ref. \cite{Renema_PRL} for
superconducting NbN bridge. Note that very similar nonlinear
dependence follows from the measurements on WSi based detector
(see inset in Fig. 10 - the data were extracted from Fig. 2 of
Ref. \cite{Baek}). We checked that for other cut-offs (we take
IDE$=0.5$ and IDE$=0.01$) the dependence $I_{thr}(E)$ is still
nonlinear (in Fig. 10 we show results for cut-off IDE$=0.5$). The
reason for the difference with Ref. \cite{Renema_PRL} is not clear
to us.

\section{Discussion}

Our experimental results correlate with preceding experiments
where the effect of magnetic field on photon and dark count rate
in SNSPD was studied. In Refs. \cite{Engel_PRB,Lusche_PRB} no
effect of low magnetic field on PCR was observed (for TaN and NbN
based detectors, respectively) while in the same range of magnetic
fields the dark count rate demonstrated strong field dependence.
Absence of change of PCR in Ref. \cite{Engel_PRB} is probably
connected with low value of the maximal magnetic field used in the
experiment ($\mu_0 H_{max}=10 mT$). In Ref. \cite{Lusche_PRB}
authors observed increase of PCR at larger magnetic fields and
found that the value of the effect depends on the wavelength - the
smaller $\lambda$ the smaller change of PCR (the similar result
was found in Ref. \cite{Korneev_MoSi} for MoSi detector). The
small {\it decrease} of PCR was observed in Ref. \cite{Lusche_PRB}
(see the bump at low magnetic fields and $I=0.78 I_{c,e}$ in Fig.
3 of Ref. \cite{Lusche_PRB}) but this effect was not discussed in
that paper.

In recent work \cite{Renema_APL} photon count rate in
superconducting NbN bridge (with width $w=150 nm$) was measured at
different magnetic fields and authors found field dependence of
PCR at $\mu_0H \gtrsim 30$ mT. From present in Fig. 4 of Ref.
\cite{Renema_APL} experimental dependence $I_c(H)$ we may conclude
that somewhere in the middle of the bridge the intrinsic defect
exists and it leads to weak change of $I_c$ at $\mu_0H \lesssim 30
mT$ (like in our NbN2 detector -  see inset in our Fig. 4, but at
different fields due to difference in the defects and widths).
This 'weak' place most probably provides finite count rate at low
currents (when IDE $\ll 1$) and PCR rapidly grows with the current
increase due to expansion of the photon-sensitive area near that
place (like near a bend in the meander). At $\mu_0H \gtrsim 30$ mT
experimental $I_c$ starts to decay with increase of magnetic field
which means that at these fields the weakest place is located at
the edge of the bridge (due to large edge screening currents
produced by applied magnetic field). It is accompanied by field
dependence of the photon count rate, qualitatively similar to our
experimental findings for large wavelengths. Unfortunately in Ref.
\cite{Renema_APL} only one wavelength ($\lambda=826$ nm) was used
and we cannot be sure in our treatment of their results. Because
in Ref. \cite{Renema_APL} PCR did not saturate at large currents
the 'crossover' current could not be observed.

In Refs. \cite{Lusche_PRB,Korneev_MoSi} authors compared experimental
 PCR(H) with prediction of the hot belt model
\cite{Bulaevskii} and find large quantitative disagreement. We
believe that the hot belt model of Ref. \cite{Bulaevskii} is not
able to explain decrease of PCR (when PCR $\simeq$ PCR$_{sat}$ and
IDE $\simeq 1$) at weak magnetic fields and stronger field
dependence of PCR for photons of smaller energies (see our
arguments in Introduction). These properties appear as inevitable
consequences of our vortex hot spot model and they are robust with
respect to the presence of edge or bulk defects which affect them
only quantitatively (they may change the position of the
'crossover' current and influence quantitatively the dependence
PCR(H) at relatively large magnetic fields).

Due to very weak field dependence of the photon count rate at low
magnetic fields $\mu_0 H\lesssim 70 mT$ for all studied
wavelengths ($\lambda =450 - 1550 nm$) we conclude that
fluctuation-activated vortex entry to the hot spot plays no role
in the photon counting for our NbN detectors. On the contrary, the
dark counts are most probably connected with fluctuation assisted
vortex nucleation in the nanowire near the 'weakest' place. It is
justified from the shift of current dependence of the dark count
rate in magnetic field which follows the change in the critical
current of the superconducting meander.

In our theoretical consideration we assume that all parts of the
meander are identical. In reality there could be variations
of the material (mean path length, critical temperature) or
geometrical (width, thickness) parameters, which of course
additionally smears dependence PCR(I) and IDE(I). But if these
inhomogenieies are small we expect they will have only
quantitative influence on the discussed here effects.

\section{Conclusion}

Experiment on the magnetic field dependence of photon count rate
in NbN based SNSPD revealed the following three main features:

(1) At low magnetic fields ($\mu_0H \lesssim 70 mT$) PCR very
weakly depends on magnetic field (for studied wavelengths $\lambda
=450 -1550$ nm), while dark count rate has pronounced field
dependence.

(2) At larger fields PCR changes with magnetic field and the
smaller the energy of the photon the stronger field dependence of
PCR.

(3) For all studied wavelengths there is
a 'crossover' current above which PCR slightly {\it decreases} while at $I<I_{cross}$ PCR {\it
increases} with increasing magnetic field. 'Crossover' current is
located close to the current at which PCR(I) saturates and reaches
plateau.

All observed features could be explained by the vortex hot spot
model:

i) In the vortex hot spot model it is assumed that the absorbed
photon creates the finite region with partially suppressed
$\Delta$ (hot spot). Appearance of such a region changes the
critical current of the nanowire. The resistive state starts at
some detection current $I_{det}$ via nucleation of the
vortex-antivortex pair inside the HS and their motion across the
nanowire if HS is located close to the center of the nanowire. If
the hot spot is located close to the edge of the nanowire the
resistive state is realized via single vortex entrance via edge
and its motion across the hot spot and nanowire.

ii) Detection current depends nonmonotonically on the position of
the hot spot across the nanowire and has maximal $I_{det}^{max}$
and minimal $I_{det}^{min}$ values. Photon count rate reaches
maximum when applied current becomes larger than $I_{det}^{max}$
and PCR gradually decreases with decreasing current due to
shrinkage of the photon-sensitive area.

iii) Perpendicular magnetic field induces the screening currents
in the nanowire and it changes $I_{det}(x)$. Detection current
decreases near the edge of the nanowire where the current density
increases and $I_{det}$ increases near the opposite edge. It leads
to the shift of $I_{det}^{min}$ to smaller currents and
$I_{det}^{max}$ to larger currents, which explains the existence
of the 'crossover' current in the experiment.

iv) When the hot spot is large (in units of coherence length) and
$\Delta$ is strongly suppressed inside HS it can strongly pin
vortices and both $I_{det}^{min}$ and $I_{det}^{max}$ slightly
varies at relatively low magnetic fields $H\lesssim H_s$. The hot
spots with small size and/or slightly suppressed $\Delta$ produces
weak pinning and detection current changes in the same magnetic
fields much stronger. It explains property (2) found in the
experiment.

v) In the detectors made in the form of meander there are right
and left bends. At small currents and weak magnetic fields these
regions are responsible for the finite photon count rate due to
locally enhanced current density. With increasing magnetic field
the whole area where the current density is locally enhanced both
in right and left bends practically does not change because in one
kind of bends current density decreases while in another it
increases. The increase of photon count rate starts only when
$I_{det}^{min}(H)$ of the straight part of meander approaches to
the transport current.

\begin{acknowledgments}

The work was partially supported by the Russian Foundation for
Basic Research (project 15-42-02365/15) and by the Ministry of
education and science of the Russian Federation (the agreement of
August 27, 2013,  N 02.Â.49.21.0003 between The Ministry of
education and science of the Russian Federation and Lobachevsky
State University of Nizhni Novgorod). The following authors
acknowledge support from the Ministry of Education and Science of
the Russian Federation: Yu.K. (unique identifier of the scientific
research RFMEFI58614x0007), A.K. (State contract
No.14.B25.31.0007), A.S. (state task No. 2327) and G.G. (state
task No. 960). A.S. also acknowledges support by grant of the
President of the Russian Federation (contract No. MK-6184.2014.2),
and G.G. acknowledges support by grant of the President of the
Russian Federation (contract No. NSh-1918.2014.2).
\end{acknowledgments}



\begin{references}
\bibitem{self} One should have in mind that plateau like dependence
of $I_c(H)$ at $H\to 0$ may exist if there is no intrinsic defect
in central part of the nanowire (see Fig. 9 for straight
nanowire). It appears when $I_c(0)\simeq I_{dep}$
\cite{Maksimova_EPL,Andrackii} due to depairing effect of
superconducting current (supervelocity). But in our detectors
$I_c(0)$ is far below $I_{dep}$ (see table 1) and this effect
could be skipped.
\end{references}
\end{document}